\documentclass[journal=nalefd, manuscript=letter]{achemso}
\setkeys{acs}{maxauthors = 0}

\usepackage{graphicx}
\usepackage{amsmath}
\usepackage{bm}
\usepackage{siunitx}
\usepackage[utf8]{inputenc}
\usepackage[english]{babel}
\usepackage{xcolor}
\usepackage{amssymb}

\newcommand*{\RU}{\mbox{$(\rightarrow\!\odot)^+$}}
\newcommand*{\LD}{\mbox{$(\leftarrow\!\otimes)^+$}}
\newcommand*{\LU}{\mbox{$(\leftarrow\!\odot)^-$}}
\newcommand*{\RD}{\mbox{$(\rightarrow\!\otimes)^-$}}
\newcommand*{\SOTunits}[1]{\mbox{\SI{#1}{mT} per \SI{e12}{A/m^2}}}

\author{T. Phuong Dao}
    \affiliation[ETHInter]
        {Laboratory for Magnetism and Interface Physics, Department of Materials, ETH Z{\"u}rich, 8093 Z{\"u}rich, Switzerland.}
    \alsoaffiliation[ETHMeso]
        {Laboratory for Mesoscopic Systems, Department of Materials, ETH Z{\"u}rich, 8093 Z{\"u}rich, Switzerland.}
    \alsoaffiliation[PSIMeso]
        {Laboratory for Multiscale Materials Experiments, Paul Scherrer Institute, 5232 Villigen PSI, Switzerland.}
    \email{phuong.dao@mat.ethz.ch}

\author{Marvin Müller}
    \affiliation[ETHFerr]
        {Laboratory for Multifunctional Ferroic Materials, Department of Materials, ETH Z{\"u}rich, 8093 Z{\"u}rich, Switzerland.}

\author{Zhaochu Luo}
    \affiliation[ETHMeso]
        {Laboratory for Mesoscopic Systems, Department of Materials, ETH Z{\"u}rich, 8093 Z{\"u}rich, Switzerland.}
    \alsoaffiliation[PSIMeso]
        {Laboratory for Multiscale Materials Experiments, Paul Scherrer Institute, 5232 Villigen PSI, Switzerland.}

\author{Manuel Baumgartner}
    \affiliation[ETHInter]
        {Laboratory for Magnetism and Interface Physics, Department of Materials, ETH Z{\"u}rich, 8093 Z{\"u}rich, Switzerland.}

\author{Ale\v{s} Hrabec}
    \affiliation[ETHMeso]
        {Laboratory for Mesoscopic Systems, Department of Materials, ETH Z{\"u}rich, 8093 Z{\"u}rich, Switzerland.}
    \alsoaffiliation[PSIMeso]
        {Laboratory for Multiscale Materials Experiments, Paul Scherrer Institute, 5232 Villigen PSI, Switzerland.}
    \alsoaffiliation[ETHInter]
        {Laboratory for Magnetism and Interface Physics, Department of Materials, ETH Z{\"u}rich, 8093 Z{\"u}rich, Switzerland.}

\author{Laura J. Heyderman}
    \affiliation[ETHMeso]
        {Laboratory for Mesoscopic Systems, Department of Materials, ETH Z{\"u}rich, 8093 Z{\"u}rich, Switzerland.}
    \alsoaffiliation[PSIMeso]
        {Laboratory for Multiscale Materials Experiments, Paul Scherrer Institute, 5232 Villigen PSI, Switzerland.}

\author{Pietro Gambardella}
    \affiliation[ETHInter]
        {Laboratory for Magnetism and Interface Physics, Department of Materials, ETH Z{\"u}rich, 8093 Z{\"u}rich, Switzerland.}
    \email{pietro.gambardella@mat.ethz.ch}

\title[short title]
  {Chiral Domain Wall Injector Driven by Spin-orbit Torques}


\begin{document}

\begin{abstract}
Memory and logic devices that encode information in magnetic domains rely on the controlled injection of domain walls to reach their full potential. In this work, we exploit the chiral coupling induced by the Dzyaloshinskii-Moriya interaction between in-plane and out-of-plane magnetized regions of a Pt/Co/AlO\textsubscript{x} trilayer in combination with current-driven spin-orbit torques to control the injection of domain walls into magnetic conduits. We demonstrate that the current-induced domain nucleation is strongly inhibited for magnetic configurations stabilized by the chiral coupling and promoted for those that have the opposite chirality. These configurations allow for efficient domain wall injection using current densities of the order of $4\times$\SI{e11}{A m^{-2}}, which are lower than those used in other injection schemes. Furthermore, by setting the orientation of the in-plane magnetization using an external field, we demonstrate the use of a chiral domain wall injector to create a controlled sequence of alternating domains in a racetrack structure driven by a steady stream of unipolar current pulses.
\end{abstract}

\section{Main Text}
The nucleation of magnetic domains underpins magnetization reversal processes and, consequently, the functioning of most types of magnetic storage devices. Domain wall (DW) racetrack memory and logic devices, in particular, require reliable control over domain nucleation and current-induced DW propagation in order to work efficiently.\cite{Allwood2005Magnetic, Parkin2008Magnetic, Alejos2017Efficient}
The problem of domain nucleation was first addressed by modifying the magnetic anisotropy of the nucleation sites using altered shapes\cite{Shigeto1999Injection,Cowburn2002Domain,Kimling2013Tuning} or ion irradiation of magnetic structures,\cite{Hyndman2001Modification, Warin2001Modification, Devolder2001Magnetic, Lavrijsen2010Controlled, Yin2017Chiral, Franken2012Shift} which favor magnetization reversal at specific locations. These methods are commonly used in field-induced domain nucleation and DW propagation studies.\cite{Boulle2008Nonadiabatic, Fernandez-Pacheco2009Domain, Laczkowski2014Experimental, Serrano-Ramon2013Modification} Current-induced domain nucleation techniques based on the Oersted field produced by a narrow write line, \cite{Hayashi2008Current} spin-transfer torque switching using magnetic tunnel junctions\cite{Ravelosona2006Domain} and using magnetization boundaries where the magnetization of the two adjacent regions are orthogonally aligned \cite{Phung2015Highly} have been shown to mitigate the shortcomings of field nucleation. These methods offer faster and more localized domain nucleation at the cost of higher device complexity.

A significant leap forward in magnetic writing was made with the advent of spin-orbit torques (SOTs),\cite{Miron2011Perpendicular, Avci2012Magnetization, Liu2012Current, Garello2013Symmetry, Kim2013Layer} which emerge at ferromagnet/heavy metal interfaces.\cite{Manchon2018Current}
Ever since the pioneering demonstration of DW propagation using SOTs,\cite{Miron2011Fast} steady advancements have led to record DW velocities,\cite{Yang2015Domain,Caretta2018Fast} higher reversal speed and reliability,\cite{Garello2014Ultrafast, Baumgartner2017Spatially, Aradhya2016SV} as well as to a deeper understanding of the DW dynamics.
In particular, it was found that, in asymmetric ferromagnet/heavy metal bilayer films, the Dzyaloshinskii-Moriya interaction (DMI), which favors orthogonal orientation of neighboring spins,\cite{Dzyaloshinsky1958thermodynamic, Moriya1960Anisotropic} plays a key role in the current-induced propagation of DWs.\cite{Thiaville2012Dynamics, Emori2013Current, Ryu2013Chiral, Boulle2013Domain, Martinez2014Current, Baumgartner2018Asymmetric} As a result of the DMI, DWs in ferromagnet/heavy metal layers with perpendicular magnetization have a chiral N\'{e}el structure, which ultimately defines the direction of the DW propagation driven by SOTs and their terminal velocity in the DW flow regime. Prior studies have shown that the chirality of DWs can be modified by tuning the strength of the DMI, magnetic anisotropy, and Zeeman interaction via external magnetic fields, which also affects the field-induced and current-induced DW depinning efficiency.\cite{Franken2014Tunable,Haazen2013Domain}
Moreover, the DMI also favors domain nucleation at the edges of magnetic stripes and dots.\cite{Baumgartner2017Spatially,Martinez2014Current,Mikuszeit2015Spin,Pizzini2014Chirality}

In this work, we demonstrate a novel mechanism to control the injection of chiral DWs in perpendicularly magnetized Pt/Co/AlO\textsubscript{x} wires, which exploits the DMI at the boundary between adjacent in-plane (IP) and out-of-plane (OOP) magnetic regions. Unlike previous investigations based on boundaries with orthogonal magnetization alignment, which have been employed for DW injection using SOTs\cite{Hayashi2012Spatial} and spin-transfer torques\cite{Phung2015Highly}, our method combines the SOTs with the chiral coupling between IP and OOP regions induced by the DMI.\cite{Luo2019Chirally} This coupling is found to strongly affect the DW nucleation process. By setting the magnetization of the IP region ($\bm{M}_{\mathrm{IP}}$) relative to the magnetization of the OOP region ($\bm{M}_{\mathrm{OOP}}$), we either enable or disable the nucleation and injection of domains in the OOP region, depending on the chirality of the IP-OOP magnetic configuration.
Once enabled, the nucleation of a domain at these boundaries requires current densities of the order of $\SI{e11}{A m^{-2}}$, which we will show is lower than for nucleation at defects or edges of magnetic stripes. Furthermore, the injection automatically disables itself after nucleation, as nucleation process changes the magnetic configuration at the boundary. This allows for further current pulses to be applied to freely propagate the injected DWs as additional injections are prevented. Therefore, chirally coupled injectors can be used to enable or disable the generation of DWs in a magnetic racetrack driven by a steady stream of clocking pulses.

\begin{figure}
\includegraphics{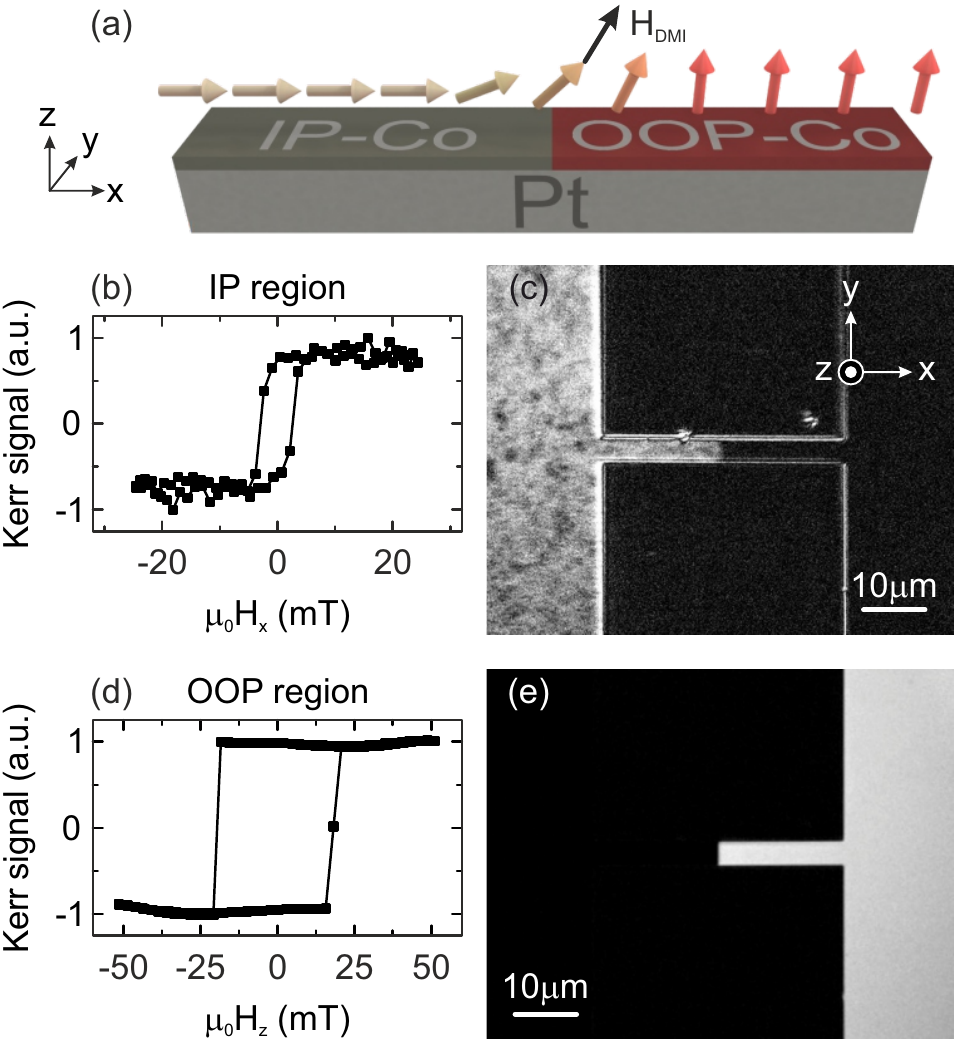}
\caption{(a) Illustration of the magnetic configuration \RU{} at an IP-OOP boundary in a Pt/Co/AlO\textsubscript{x} wire. (b) In-plane hysteresis loop and (c) corresponding differential MOKE image recorded with a MOKE microscope in longitudinal mode. (d) Out-of-plane hysteresis loop and (e) corresponding differential MOKE image recorded with a MOKE microscope in polar mode. The IP-OOP boundary is located at the border between the black and white regions. Here, the magnetization is strongly influenced by the DMI, which favors the \RU{} or \LD{} configurations.}
\label{fig:fig1}
\end{figure}

In Figure \ref{fig:fig1}a, we show the basic structure of the chiral DW injector, namely a Pt/Co/AlO\textsubscript{x} wire consisting of two regions with IP and OOP magnetization, respectively. At the IP-OOP boundary between the two regions, the magnetic configuration is determined by the interplay between the exchange interaction, magnetic anisotropy, and DMI. The effect of the DMI can be described by an effective field, $\bm{H}_{\mathrm{DMI}}$, which acts on the local magnetization direction $\bm{M}$. Considering for simplicity a one-dimensional wire elongated along $\bm{x}$, the effective DMI field is given by:\cite{Thiaville2012Dynamics,Rohart2013Skyrmion}
\begin{equation}
\bm{H}_{\mathrm{DMI}}=\dfrac{2D}{\mu_\mathrm{0} M_\mathrm{s}}
\begin{pmatrix}
-\dfrac{dm_\mathrm{z}}{dx},& 0, &\dfrac{dm_\mathrm{x}}{dx}
\end{pmatrix},
\label{eq:Hdmi}
\end{equation}
where $D$ is the material-dependent DMI constant in units of $\si{J m^{-2}}$, $\mu_\mathrm{0}$ is the vacuum permeability, $M_\mathrm{s}$ the saturation magnetization, and $m_\mathrm{x}$ and $m_\mathrm{z}$ are the components of the normalized magnetization vector ${\bm{m}=\bm{M}/M_\mathrm{s}}$. The sign of $D$ determines the favored chirality of the IP-OOP boundary, namely the sense of rotation of $\bm{m}$ in the $xz$-plane. In Pt/Co/AlO\textsubscript{x}, $D$ is negative, which corresponds to a counterclockwise chirality.\cite{Belmeguenai2015Interfacial}
Since $\bm{M}_{\mathrm{IP}}$ can point along $+\bm{x}$ $(\rightarrow)$ or $-\bm{x}$ $(\leftarrow)$ and $\bm{M}_{\mathrm{OOP}}$ can point along $+\bm{z}$ $(\odot)$ or $-\bm{z}$ $(\otimes)$, we can identify four distinct configurations in our devices.
These configurations differ in the energy density
\begin{equation}
E_{\mathrm{DMI}}=-\bm{M}\cdot\bm{H}_{\mathrm{DMI}} = -2D \left(-\dfrac{dm_\mathrm{z}}{dx}m_\mathrm{x}+\dfrac{dm_\mathrm{x}}{dx}m_\mathrm{z} \right)
\label{eq:Edmi}
\end{equation}
integrated over the direction perpendicular to the IP-OOP boundary. The configurations \RU{}, illustrated in Figure \ref{fig:fig1}a, and \LD{} have a low energy and are stabilized by the chiral coupling, which is denoted by "$^+$", whereas the \RD{} and \LU{} configurations are destabilized by the chiral coupling, which is denoted by "$^-$".

Taking $D= \SI{-1.2}{mJ m^{-2}}$,\cite{Belmeguenai2015Interfacial,Luo2019Chirally} the magnitude of $\left\lVert\mu_{0}\bm{H}_{\mathrm{DMI}}\right\rVert$ can be estimated by employing micromagnetic simulations to determine the profile of the magnetization across the IP-OOP boundary, as required by Equation \ref{eq:Hdmi}.
Using this method, we estimate an average effective field $\left\lVert\mu_{0}\bm{H}_{\mathrm{DMI}}\right\rVert\approx\SI{100}{mT}$ for Pt/Co/AlO\textsubscript{x}. In the absence of external magnetic fields, $\bm{H}_{\mathrm{DMI}}$ can be strong enough to revert the unstable configurations \RD{} and \LU{} back into the stable configurations \RU{} or \LD{}, which is generally the case in nanoscopic islands.\cite{Luo2019Chirally} \RD{} and \LU{} can be made metastable by increasing the dimension $l$ perpendicular to the boundary, which in turn increases the energy barrier for magnetization reversal. For $l\gtrsim{}\SI{100}{nm}$ in Pt/Co/AlO\textsubscript{x}, the energy barrier is determined by the energy required to nucleate a new DW. 
In this work, we study the SOT-induced DW injection at IP-OOP boundaries and make use of the difference in energy between the stable and metastable configurations to enable and disable the current-induced injection.

The fabrication of IP-OOP boundaries requires precise local control over the magnetic anisotropy. Spatial engineering of the magnetic anisotropy has previously been achieved using ion irradiation\cite{Hyndman2001Modification, Warin2001Modification, Devolder2001Magnetic, Lavrijsen2010Controlled,Franken2014Tunable,Haazen2013Domain,Hayashi2012Spatial}
 and electric gating. \cite{Weisheit2007Electric,Maruyama2009Large,Matsukura2015Control,Schott2017Skyrmion}
In this work, we utilize selective oxidation in order to modify the magnetic anisotropy of Pt/Co/(Al)/AlO\textsubscript{x} layers. The under-oxidized Pt/Co/Al/AlO\textsubscript{x} system is known to exhibit IP anisotropy, whereas the oxidized Pt/Co/AlO\textsubscript{x} system possesses perpendicular magnetic anisotropy, induced by the formation of Co-O bonds at the Co/AlO\textsubscript{x} interface.\cite{Monso2002Crossover,Rodmacq2003Crossovers,Manchon2008Analysis,Lacour2007Magnetic,Schott2017Skyrmion,Dieny2017Perpendicular,Rau2014Reaching} Compared to ion irradiation, oxidation provides a way to change the anisotropy by only modifying the top interface of the magnetic layer leaving the magnetic material pristine. By locally tuning the oxidation, we can freely define regions with different anisotropy. We achieve this selective oxidation with patterned masks made by electron beam lithography, which has a lateral resolution at the nanometer scale, and subsequent oxygen plasma. 
The width of the IP-OOP boundary is determined by the sharpness of the mask used for the oxidation. For overhanging resist masks, we estimate that shadowing during the oxidation process leads to a maximal width of \SI{20}{nm}.

Using this method, we fabricated a magnetic wire consisting of two regions with different anisotropies, as shown in Figure \ref{fig:fig1}a, where the left half is IP and the right half is OOP, and we examined these wires using a home-built wide field magneto-optical Kerr effect (MOKE) microscope.
The longitudinal MOKE measurements, which are sensitive to the $m_\mathrm{x}$ component, are shown in \mbox{Figure \ref{fig:fig1}b and \ref{fig:fig1}c}.
The IP region has uniaxial anisotropy that favors the direction parallel to the wire axis ($\pm\bm{x}$), as shown by the almost square hysteresis curve in Figure \ref{fig:fig1}b measured as a function of in-plane magnetic field $H_\mathrm{x}$.
The differential MOKE image (Figure \ref{fig:fig1}c) was obtained by taking the difference between the images of the remanent states taken at zero field after saturating along $+\bm{x}$ and $-\bm{x}$, respectively.
The contrast in the image indicates that the IP signal originates from the left half of the wire.
The uneven contrast in the IP region is a result of the region breaking into domains and the low signal to noise ratio of the longitudinal MOKE.
To measure the OOP component of the magnetization, $m_\mathrm{z}$, we switch the microscope to the polar mode and repeat the hysteresis measurement but now with a magnetic field $H_\mathrm{z}$ along $\pm\bm{z}$. The square hysteresis loop, shown in Figure \ref{fig:fig1}d, implies a uniaxial OOP anisotropy and, from the differential MOKE image in Figure \ref{fig:fig1}e, we can confirm that the signal is coming from the right half of the wire.
Furthermore, this observation confirms that the IP region is fully IP, as small OOP domains would be clearly visible due to strong contrast in the polar MOKE.
This demonstrates that our fabrication method succeeds in sharply defining regions of different anisotropies.

\begin{figure}
\includegraphics{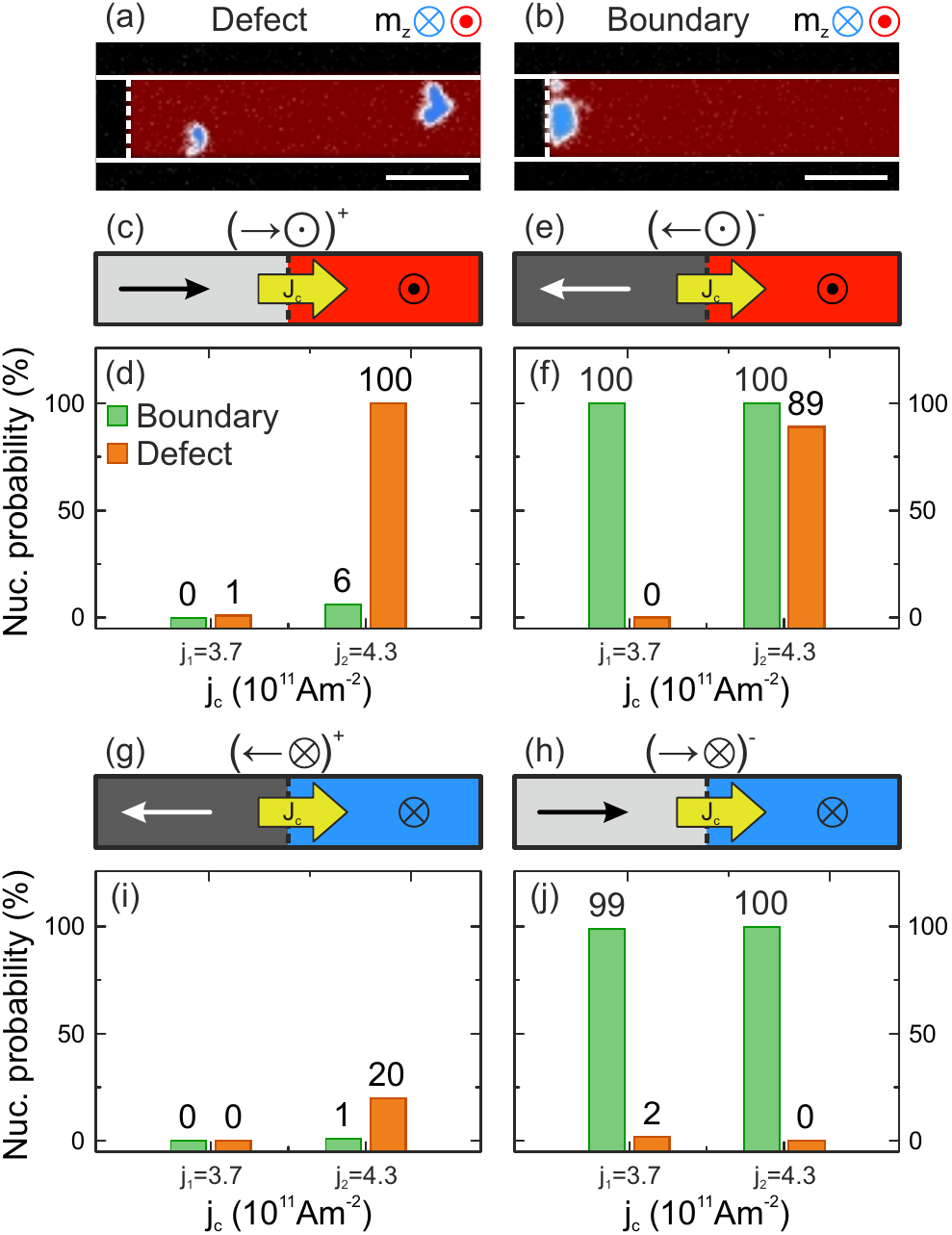}
\caption{Differential MOKE images of wires with IP-OOP boundaries showing current-induced domain nucleation at defect sites (a) or at the boundary itself (b). The scale bars correspond to \SI{4}{\mu m}. (c) Illustration of the magnetic configuration \RU{} before the indicated current pulse $j_\mathrm{c}$ was applied. (d) The nucleation probability at defects (orange) and at the boundary (green) for \RU{}, measured for two different current densities. (e-j) Remaining three configurations and associated nucleation probabilities. For \RU{} and \LD{}, the nucleation at the boundary is strongly suppressed, whereas for \LU{} and \RD{} it is promoted. The nucleation at defects displays no dependence on the chirality of the magnetic configurations.}
\label{fig:fig2}
\end{figure}

To investigate the current-induced domain nucleation in the OOP region we used the MOKE microscope in polar mode to locate the domain nucleation sites in a \SI{4}{\micro\meter} wide magnetic wire. Examples of the differential MOKE images, which show the difference between an image taken before and after nucleation attempts, are shown in Figure \ref{fig:fig2}a and \ref{fig:fig2}b. As can be seen, the nucleation generally occurs not only at the IP-OOP boundary but also elsewhere on the sample due to random defects in the film. We elucidate the difference between the random thermal nucleation at defects and the deterministic nucleation at the IP-OOP boundary by probing both the current dependence and field dependence of the domain nucleation.

We begin with the domain nucleation in the \RU{} configuration as shown in Figure \ref{fig:fig2}c. This configuration was set with two short, external magnetic field pulses, first $\mu_{0}H_\mathrm{z}=\SI{100}{mT}$ in $+z$ direction and then $\mu_{0}H_\mathrm{x}=\SI{50}{mT}$ in $+x$ direction. We then sent the current pulse in $+x$ direction to nucleate a $\otimes$ domain, with no applied magnetic field, and compared the images before and after the current pulse. This procedure was repeated one hundred times for two current densities, namely $j_\mathrm{1}=\SI{3.7e11}{A m^{-2}}$ and $j_\mathrm{2}= \SI{4.3e11}{A m^{-2}}$. The statistics of the domain nucleation can be found in Figure \ref{fig:fig2}d, where the green bars represent the nucleation that occurred at the boundary. For $j_\mathrm{1}$, we observe almost no nucleation in the wire, neither at the IP-OOP boundary nor at defects. For the higher current density $j_\mathrm{2}$, the nucleation at the boundary increases slightly to $\SI{6}{\%}$ while defect mediated nucleation increases to $\SI{100}{\%}$.
So far, the data indicate that domain nucleation at the IP-OOP boundary is much less likely than nucleation at defects.

We then repeated the nucleation experiment for \LU{} as illustrated in Figure \ref{fig:fig2}e. We now saturate with $H_\mathrm{z}$ and $-H_\mathrm{x}$ field pulses, while keeping the same current direction. For this configuration, unlike for \RU{}, the nucleation at the IP-OOP boundary was guaranteed even with the lower current density $j_\mathrm{1}$ as shown in Figure \ref{fig:fig2}f. The defect mediated nucleation on the other hand, is still rare. For $j_\mathrm{2}$, the boundary mediated nucleation probability stays at $\SI{100}{\%}$ while the nucleation probability at defects increases dramatically, as was observed in the \RU{} configuration.
The nucleation probability of the remaining two configurations \LD{} and \RD{}, see Figure \ref{fig:fig2}g and \ref{fig:fig2}h, respectively, confirms the asymmetry.
The juxtaposition of the boundary mediated nucleation probabilities of stabilized and destabilized configurations reveals the role of the DMI in suppressing or promoting the domain nucleation in the OOP region, especially when we take into account how little the nucleation at defects changes between the different chiralities. Furthermore, the boundary mediated nucleation consistently requires less current in the destabilized states \LU{} and \RD{} compared to the nucleation at defects or at the edge of the wire.

The asymmetric domain nucleation probabilities of \RU{} and \LU{} can be explained by considering the action of the effective field $\bm{H}_{\mathrm{DMI}}$, and more specifically of its $z$-component. As stated in Equation \ref{eq:Hdmi}, the $z$-component of $\bm{H}_{\mathrm{DMI}}$ depends on the change of $m_\mathrm{x}$ and $D$, which is negative for our system. Evidently, $m_\mathrm{x}$ always goes to zero in the OOP region whether it is $\odot$ or $\otimes$, so the $z$-component of $\bm{H}_{\mathrm{DMI}}$ is entirely determined by the magnetization direction of the IP region.
For \RU{}, where $\bm{M}_{\mathrm{IP}}$ points along $+\bm{x}$, the $z$-components of both $\bm{H}_{\mathrm{DMI}}$ and $\bm{M}_{\mathrm{}\mathrm{OOP}}$ are positive. As a consequence, the DMI opposes the reversal and strongly inhibits domain nucleation.
When $\bm{M}_{\mathrm{IP}}$ is reversed, as in the \LU{} configuration, $\bm{H}_{\mathrm{DMI}}$ is also reversed and has a negative $z$-component opposing $\bm{M}_{\mathrm{OOP}}$, which destabilizes this configuration and greatly facilitates the reversal of $\bm{M}_{\mathrm{OOP}}$ using SOTs. The domain nucleation in the remaining configurations \LD{} and \RD{} follows the same logic, where $\bm{H}_{\mathrm{DMI}}$ inhibits and promotes the nucleation, respectively.

Finally, we note that the boundary mediated nucleation in the OOP region was only observed for a current flowing along $+\bm{x}$, i.e., from the IP to the OOP region. This is a natural consequence of the DW propagation direction being parallel to the injected current in Pt/Co/AlO\textsubscript{x}.\cite{Miron2011Fast}
For the opposite current direction, the nucleation should occur in the IP region, but this was not observed. We ascribe the absence of nucleation in the IP region to the combination of the uniaxial anisotropy along $\bm{x}$ and the demagnetizing field, which stabilizes the IP magnetization, as well as to the fact that SOTs are most effective in switching the magnetization when the current and magnetization are perpendicular rather than parallel to each other.

\begin{figure}
\includegraphics{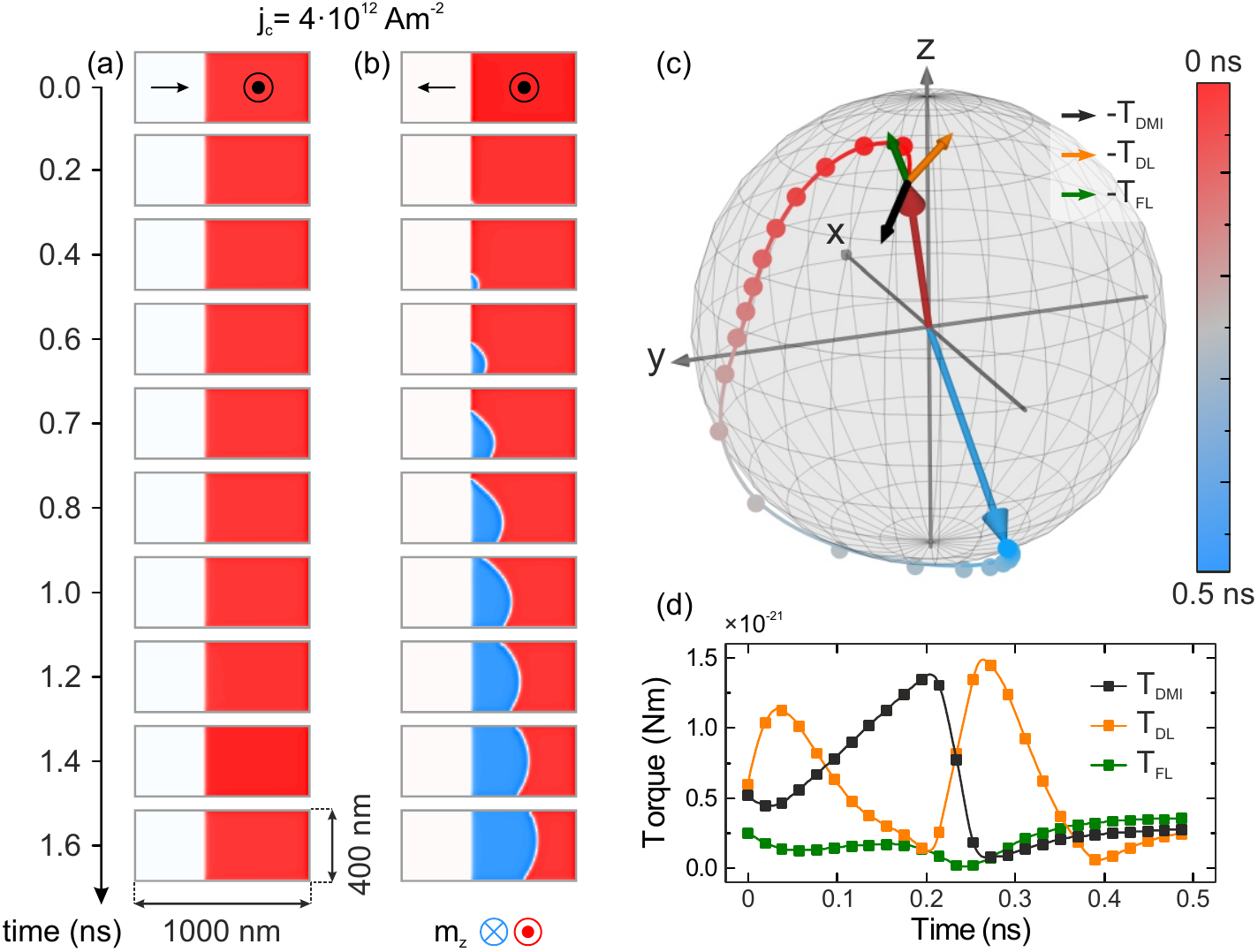}
\caption{Micromagnetic simulation of the domain nucleation at IP-OOP boundaries for two different magnetic configurations. For a current density of $j_\mathrm{c}=\SI{4e12}{A m^{-2}}$, no nucleation is observed in (a) for the \RU{} configuration. For the \LU{} configuration in (b), however, the nucleation is facilitated and a domain is injected into the wire. This leads to a full reversal of the OOP region resulting in a stable \LD{} configuration. (c) Trajectory of the magnetization extracted from OOMMF simulations at the nucleation site, which to a large part is the result of the combined action of the SOTs and DMI. Combining this with the temporal evolution of the torques in (d), we can deduce that the nucleation is induced by SOTs in the first \SI{0.1}{ns}, but that the main driving force for the reversal is the chiral coupling induced by the DMI. Once the $z$-component of the magnetization changes sign, the magnetic configuration follows the chirality imposed by the DMI and the torque exerted by the DMI quickly falls off.}
\label{fig:fig3}
\end{figure}

To shed more light on the asymmetric domain nucleation and the influence of the DMI on the configuration at the IP-OOP boundary, we performed micromagnetic simulations using OOMMF\cite{Donahue1999OOMMF} (Figure 3).
The simulated sample consists of a square \SI{400}{\nano\meter} $\times$ \SI{400}{\nano\meter} IP region, which was kept small to save computing time, and a \SI{400}{nm} $\times$ \SI{600}{\nano\meter} OOP region with a sharp IP-OOP boundary. The unixial anisotropy of the OOP (IP) region was set to $K_{\mathrm{OOP}} = \SI{625}{kJ m^{-3}}$ ($K_{\mathrm{IP}} = \SI{650}{kJ m^{-3}}$). Note that, in our convention, $K_{\mathrm{IP}}$ is positive, reflecting the IP uniaxial anisotropy along the wire axis found in the longitudinal MOKE measurements.
In order to simplify the model and reduce the parameter space, other interface-dependent quantities, such as the DMI constant $D = \SI{-1.2}{mJ m^{-2}}$, field-like torque $T_{\mathrm{FL}} = \SOTunits{7}$, and damping-like torque $T_{\mathrm{DL}} = \SOTunits{18}$,\cite{Baumgartner2017Spatially,Baumgartner2018Asymmetric} are assumed to be constant across the IP-OOP boundary. In general, however, these parameters may vary with the oxidation profile. Similarly, we assumed the same bulk material parameters for the two regions, namely $M_{\mathrm{s}} = \SI{900}{kA m^{-1}}$ and exchange coupling $A_{\mathrm{ex}} = \SI{11e-11}{J m^{-1}}$.
In Figure \ref{fig:fig3}a and b, we show the results of the simulations of the nucleation and propagation processes for the \RU{} and \LU{} configurations, respectively. In both simulations, a current density of \SI{4e12}{A m^{-2}} was applied and no external magnetic field was present. In Figure \ref{fig:fig3}a, the first frame at \SI{0}{ns} represents the relaxed configuration \RU{} with no current applied. In agreement with our measurements, we observe no switching in this case. We do however observe a small tilt of $\bm{M}_{\mathrm{OOP}}$ towards the $y$-direction due to the SOTs, but the effective DMI field has a positive $z$-component opposing magnetization reversal. In contrast, for the \LU{} configuration shown in Figure \ref{fig:fig3}b, a $\otimes$ domain nucleates in the OOP region after \SI{0.25}{ns} as $\bm{H}_{\mathrm{DMI}}$ now assists the magnetization reversal. The nucleated domain quickly grows until the DW eventually spans the whole width of the wire at \SI{1.2}{ns} and then continues to propagate in the direction of the current. Interestingly, the nucleation always occurs at the bottom edge of the wire for this configuration.
This behavior is the result of the canting of the magnetization at the edge of the sample due to the DMI.\cite{Baumgartner2017Spatially} For the \LU{} configuration, the initial tilt of the magnetization at the bottom edge favors its rotation in the same sense as that promoted by the SOTs, whereas at the top edge the tilt is in the opposite direction, inhibiting the effect of the SOTs. In the other metastable configuration \RD{}, the canting is reversed and nucleation always starts at the top edge.

To gain deeper insight into the dynamics of the nucleation process, we analyze the trajectory of $\bm{M}_{\mathrm{OOP}}$ in a single cell of the OOMMF simulation, located at the bottom edge of the wire, \SI{8}{nm} away from the IP-OOP boundary.
In Figure \ref{fig:fig3}c, we show the directions of the torques that act on $\bm{M}_{\mathrm{OOP}}$ at the start of the reversal, i.e., the damping-like torque, $\bm{T}_{\mathrm{DL}}\propto\bm{M}\times\left(\bm{y}\times\bm{M}\right)$, field-like torque, $\bm{T}_{\mathrm{FL}}\propto\bm{y}\times\bm{M}$ and $\bm{T}_{\mathrm{DMI}}$ exerted by the chiral coupling. In the resulting trajectory of $\bm{M}_{\mathrm{OOP}}$ in Figure \ref{fig:fig3}c, it can be seen how the magnetization first tilts toward the $z$-axis and then rapidly reverses by tilting towards the $y$-direction. The temporal evolution of the torques, presented in Figure \ref{fig:fig3}d, demonstrates that the initial tilt is induced by the SOTs but, once $\bm{M}_{\mathrm{OOP}}$ has moved sufficiently away from its metastable position, the effect of the DMI rapidly increases, eventually pulling the magnetization towards $-\bm{z}$. At the same time, the SOTs decrease, showing that the DMI becomes the main driver of magnetization reversal, whereas the SOTs are responsible for starting the process. This is in agreement with the experimentally observed reduction of the critical current density for the nucleation at the IP-OOP boundary, which is mainly just required to start the reversal.

In our experiments, the boundary mediated nucleation in the \SI{4}{\micro\meter} wide wires shown in Figure \ref{fig:fig2}a was generally more heterogeneous compared to the nucleation observed in the simulations. In particular, bubble-like domains formed at the boundary and merged with further application of current pulses. We ascribe this behavior to the quality of the IP-OOP boundary. 
If, instead of a perfectly straight boundary, we consider a boundary with a notch, a small OOP area next to the notch will be associated with a longer IP-OOP boundary, which increases the chiral coupling and, as a consequence, the nucleation efficiency in that area.
For narrower lines we generally observed more homogeneous nucleation and a stronger asymmetry in the nucleation probability.

\begin{figure}
\includegraphics{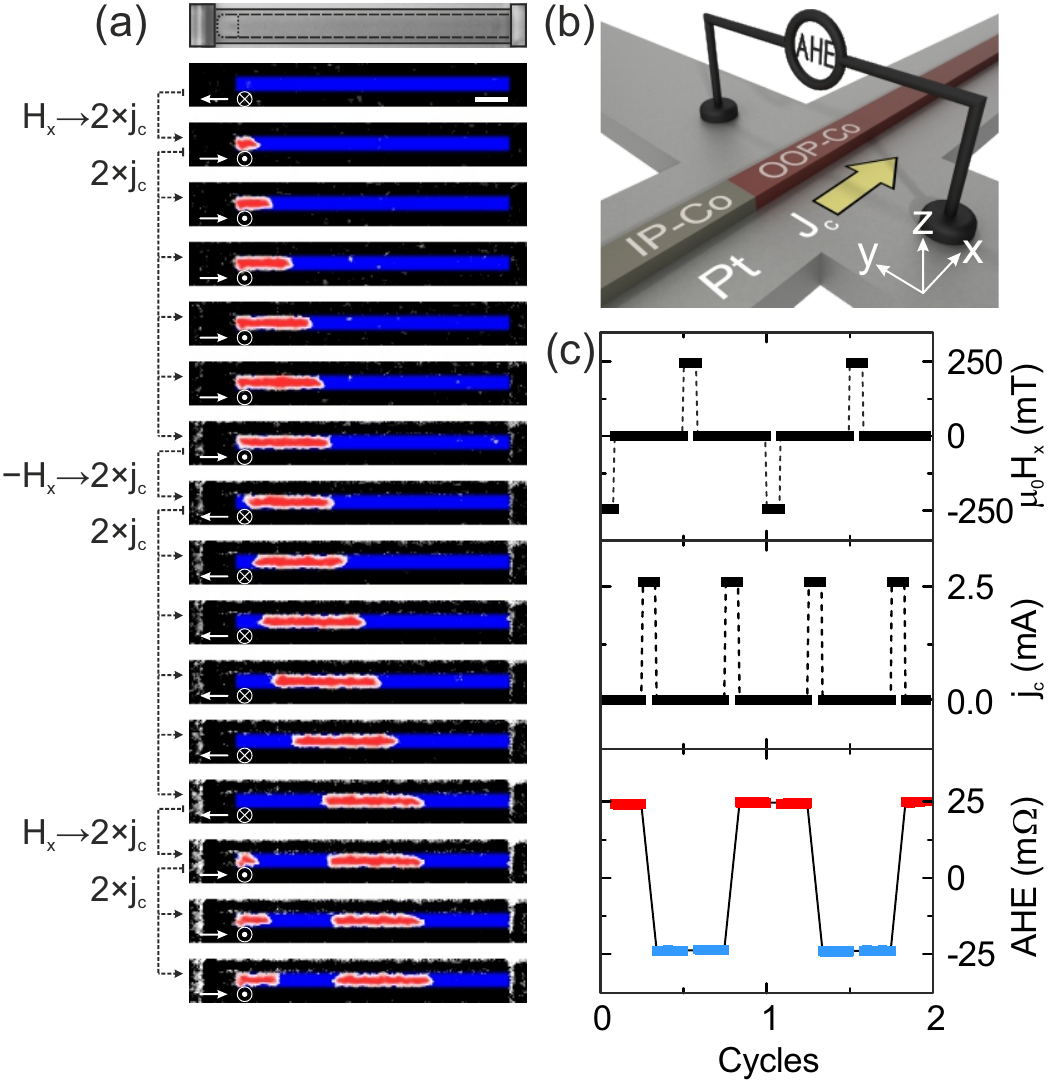}
\caption{(a) Optical microscope image of an \SI{800}{nm} wide DW conduit (top) and differential polar MOKE images of domain injection in the conduit. Starting from a saturated state, we make use of the asymmetric domain nucleation probabilities to inject a controlled sequence of alternating $\odot$ and $\otimes$ domains.
The external magnetic field, $\left\vert \mu_{0}H_\mathrm{x}\right\vert = \SI{80}{mT}$, and the two current pulses, $j_\mathrm{c} = \SI{7e11}{A m^{-2}}$, applied between each step are indicated to the left of the images.
(b) Illustration of a \SI{100}{nm} wide DW conduit on top a Pt Hall cross. In this device, the injection can be measured electrically using the anomalous Hall effect. The top two plots in (c) are the field and current protocols, where each peak represents one corresponding pulse, used to achieve the switching presented in the bottom plot. After each current pulse, the AHE changes sign indicating that $\bm{M}_{\mathrm{OOP}}$ reversed. The domain nucleation in the OOP region is enabled by reversing the IP region with an external field $H_\mathrm{x}$ between each current pulse. The scale bar corresponds to \SI{2}{\mu m}.}
\label{fig:fig4}
\end{figure}

As the next step, we utilize the asymmetric domain nucleation probabilities to deterministically inject and propagate DWs into a \SI{800}{nm} wide conduit (see Figure \ref{fig:fig4}a) using a stream of current pulses of fixed amplitude and direction.
We used a relatively high current density of $j_\mathrm{c} = \SI{7e11}{A m^{-2}}$, which is not needed for the nucleation but to overcome the pinning for DW propagation.
Since the higher $j_\mathrm{c}$ also increases the DW velocity, we shortened the pulse length to \SI{35}{ns} in order to maintain a fine level of control over the propagation.
The first differential MOKE image shown in Figure \ref{fig:fig4}a corresponds to the initial magnetization configuration \LD{} of the DW conduit, which we set with two consecutive external magnetic field pulses, $-\mu_{0}H_\mathrm{z}=\SI{-80}{mT}$ followed by $-\mu_{0}H_\mathrm{x}=\SI{-80}{mT}$. As this is a stable configuration, we first prime the DW injection by reversing $\bm{M}_{\mathrm{IP}}$ with an external field pulse $H_\mathrm{x}$. After changing the configuration to \RD{}, the first current pulse injects an $\odot$ domain into the conduit, as shown in the second frame in Figure \ref{fig:fig4}a. Every subsequent current pulse only causes the DW to propagate since the configuration was changed to \RU{} after the injection, which is stable. The somewhat uneven DW velocity between the frames is caused by the pinning in Pt/Co/AlO\textsubscript{x}. Nevertheless, the length of the injected domain can be precisely controlled and, once the desired domain size is reached, the nucleation can be once again enabled by reversing $\bm{M}_{\mathrm{IP}}$ with an external field pulse $-H_\mathrm{x}$ to give the \LU{} configuration. After the injection of another $\otimes$ domain of arbitrary length, we cycled back to the starting configuration \LD{}. 
This whole process can be repeated to obtain sequences of domains with arbitrary domain lengths.

To further demonstrate the injection process in miniaturized structures, we fabricated DW conduits with a width of \SI{100}{nm} and used the same nucleation protocol but with longer pulses, namely \SI{100}{ms}, to reverse the entire length of the wire (\SI{10}{\mu m}).
Since the width is below the optical resolution of MOKE, the conduits were fabricated on a Hall cross to measure $\bm{M}_{\mathrm{OOP}}$ electrically using the anomalous Hall effect (AHE), as shown in Figure \ref{fig:fig4}b.
As the AHE is proportional to $m_\mathrm{z}$, an inversion of the Hall signal signifies the passage of a DW and the subsequent reversal of the magnetization of the wire.
In Figure \ref{fig:fig4}c, we demonstrate how two cycles of switching, from $\odot$ to $\otimes$ and back again, are achieved using the asymmetric domain nucleation probabilities.
Each cycle starts with a high AHE signal for the configuration \RU{}. Nucleation is then enabled by applying $-H_\mathrm{x}$. The AHE remains positive indicating that $\bm{M}_{\mathrm{OOP}}$ has not yet changed but, once a current $j_\mathrm{c}$ is applied, $\bm{M}_{\mathrm{OOP}}$ reverses, causing a sharp drop in the AHE. To switch the magnetization back, the same procedure is followed but this time applying $+H_\mathrm{x}$ instead of $-H_\mathrm{x}$ to go from \LD{} to \RD{}.

In conclusion, we exploit the chiral coupling induced by the DMI at the boundary between regions with IP and OOP magnetic anisotropy to achieve current-induced injection of chiral DWs in Pt/Co/AlO\textsubscript{x} wires. These IP-OOP boundaries were created by engineering the magnetic anisotropy Pt/Co/AlO\textsubscript{x} using selective oxidation of the IP magnetized regions to induce perpendicular magnetic anisotropy.
We found that the magnetization configuration and dynamics at the IP-OOP boundary are strongly influenced by the chiral coupling, which we can express as an average effective field $\left\lVert\mu_{0}\bm{H}_{\mathrm{DMI}}\right\rVert\approx\SI{100}{mT}$ that stabilizes configurations that follow the chirality imposed by the DMI and destabilizes configurations with the opposite chirality.
In the studied systems, in contrast to Ref. \cite{Luo2019Chirally}, this effective field is not strong enough to revert destabilized configurations back to stable ones, as the IP and OOP uniaxial anisotropy in the two respective regions balance the DMI, making these configurations metastable.
For current-induced nucleation, the consequence of $\bm{H}_{\mathrm{DMI}}$ is that switching from metastable to stable configurations is greatly facilitated, lowering the required current density, whereas switching from stable to metastable configurations is strongly inhibited. Micromagnetic simulations show that current-induced nucleation at these IP-OOP boundaries is initiated by SOTs, but that the main driving force for the magnetization reversal is $\bm{H}_{\mathrm{DMI}}$. This is in agreement with the lower critical current densities systematically observed for domain nucleation at the destabilized boundary compared to nucleation at defects or at the edges of the device.

Finally, we explore the possibility of integrating these IP-OOP boundaries into DW conduits of different widths in order to exploit the chiral coupling for DW injection. Using a simple field and current pulse scheme, we successfully demonstrate the injection of an alternating sequence of $\odot$ and $\otimes$ domains into conduits as narrow as 100~nm. The inherent asymmetry in the nucleation probability for different chiralities provides a means to freely configure the length of the injected domains by controlling the magnetization of the IP region.
Our devices have an inherent one-dimensional design with only two contacts, making their implementation simpler than Oersted-field or magnetic tunnel junctions. The injector is also insensitive to OOP magnetic fields, making it very flexible for DW studies. If required, the need for an external IP field to set the initial state of the injector can be overcome by adopting one of the following solutions: i) switching of the IP magnetization by SOTs using an IP region patterned at an angle with respect to the current direction (two-terminal device), ii) switching of the IP magnetization by SOTs using a four-terminal geometry, and iii) using a Y-shaped injector with two IP regions with opposite magnetization and toggling the current between the two arms of the Y (three-terminal device).
Compared to DW injectors based on spin-transfer torques at IP-OOP boundaries in the absence of DMI,\cite{Phung2015Highly} the application of SOTs in combination with the chiral coupling lowers the critical current density for deterministic injection in nanoscale devices by an order of magnitude, down to $\SI{e11}{A m^{-2}}$ without putting an upper limit on the current density to reach higher DW velocities. Moreover, the asymmetric domain nucleation probability using chirally coupled IP-OOP boundaries gives complete control over the length of the injected domains using a stream of unipolar current pulses with a single current density and pulse length.

\begin{acknowledgement}
We acknowledge the Swiss National Science Foundation for financial support through grant No. 200020-172775. A.H. was funded by the European Union’s Horizon 2020 research and innovation programme under Marie Skłodowska-Curie grant agreement number 794207 (ASIQS).
\end{acknowledgement}



\bibliography{main}

\end{document}